\documentclass[aps,numerical,graphicx,showpacs,preprint,prl]{revtex4-1}
\usepackage{amsmath}
\usepackage{graphicx}

\begin{document}

\def\Journal#1#2#3#4{{#1 }{\bf #2, }{ #3 }{ (#4)}}

\def\BiJ{ Biophys. J.}
\def\Bios{ Biosensors and Bioelectronics}
\def\LNC{ Lett. Nuovo Cimento}
\def\JCP{ J. Chem. Phys.}
\def\JAP{ J. Appl. Phys.}
\def\JMB{ J. Mol. Biol.}
\def\JPC{ J. Phys: Condens. Matter}
\def\CMP{ Comm. Math. Phys.}
\def\LMP{ Lett. Math. Phys.}
\def\NLE{{ Nature Lett.}}
\def\NPB{{ Nucl. Phys.} B}
\def\PLA{{ Phys. Lett.}  A}
\def\PLB{{ Phys. Lett.}  B}
\def\PNAS{Proc. Natl. Am. Soc.}
\def\PRL{ Phys. Rev. Lett.}
\def\PRA{{ Phys. Rev.} A}
\def\PRE{{ Phys. Rev.} E}
\def\PRB{{ Phys. Rev.} B}
\def\PD{{ Physica} D}
\def\ZPC{{ Z. Phys.} C}
\def\RMP{ Rev. Mod. Phys.}
\def\EPJD{{ Eur. Phys. J.} D}
\def\SAB{ Sens. Act. B}
\title{
The Puzzling of Stefan-Boltzmann Law: Classical or Quantum Physics
}
\author{Lino Reggiani}
\email{lino.reggiani@unisalento.it}
\affiliation{Dipartimento di Matematica e Fisica, ``Ennio de Giorgi'',
Universit\`a del Salento, via Monteroni, I-73100 Lecce, Italy}

\author{Eleonora Alfinito}
\affiliation{Dipartimento di Matematica e Fisica, ``Ennio de Giorgi'',
Universit\`a del Salento, via Monteroni, I-73100 Lecce, Italy}

\
\date{\today}
\begin{abstract}
Stefan-Boltzmann law, stating the fourth power temperature dependence of the radiation emission by a black-body, was empirically deduced  by Stefan in 1874  
by fitting existing experiments and theoretically validated by Boltzmann in 1884 on the basis of a classical physical model involving thermodynamics principles and 
the radiation pressure predicted by Maxwell equations. At first sight the electromagnetic 
(EM) gas assumed by Boltzmann and identifiable as an ensemble of $N$ classical normal-modes, looks like an extension of the classical model of the massive ideal-gas. 
Accordingly, for this EM gas the internal total energy, $U$, was taken to be function of volume $V$ and temperature $T$ as $U=U(V,T)$, and the equation of state was given by
$U=3PV$, with $P$ the radiation pressure.  In addition, Boltzmann implicitly assumed that, for given values of $V$ and $T$, $U$ and $N$ would take finite values. 
However, from one hand these assumptions are not justified by Maxwell equations since, in vacuum (i.e. far from the EM sources), according to classical statistics, the values of $U$and $N$ diverge, the so called ultraviolet catastrophe.
From another hand, Boltzmann derivation of Stefan law is found to be macroscopically compatible with its derivation from quantum statistics announced by Planck in 1901 \cite{planck01}.
Accordingly, this letter presents a solution of this puzzling classical/quantum compatibility by noticing that the implicit assumption made by Boltzmann is fully justified by quantum statistics. Furthermore, we shed new light on the interpretation of recent classical simulations of a black-body carried out  by Wang, Casati, and Benenti in 2022 who found an analogous puzzling compatibility to induce  speculations on classical physics and black-body radiation that  are claimed to require a critical reconsideration of the role of classical physics for the understanding of quantum mechanics.

\end{abstract}
%

\maketitle 

\section{Introduction and Boltzmann model} 
The Stefan Boltzmann (SB) law for the radiation emission of a black-body was empirically deduced  by Stefan in 1874 \cite{stefan} in a first attempt to fit 
existing experiments and
theoretically deduced by Boltzmann in 1884 \cite{boltzmann,bartoli1884} on the basis of a physical model combing the  first and second principle of thermodynamic and 
the law for the radiation pressure predicted by Maxwell equations. 
At first sight, the electromagnetic (EM) gas assumed by Boltzmann looks like an extension of the classical model of a massive ideal gas where, 
following Maxwell equations: (i) the internal energy is function of volume and temperature as: 
\begin{equation}
U=U(V,T)
\end{equation}
and, (ii) the equation of state is given by
\begin{equation}
U=3 PV
\end{equation}
with $V$ and $T$, respectively, the black-body volume and the absolute temperature of the radiation emitted by its surface that should be considered as the corresponding thermal reservoir, and $P$ the pressure exerted by the EM gas on the walls of the black-body \cite{leff02,crepeau09}.  
A further  implicit assumption made by Boltzmann was that, for finite values of  $V$ 
and $T$, also $U$ takes finite values.
However, this assumption is not compatible for a black-body filled by a classical EM gas that, by following classical statistics, is described by an ensemble with an infinite number of classical normal-modes,  each with energy given by
\begin{equation}
\epsilon_{cm}=pc
\end{equation}
where the $cm$ subscript stands for classical-modes,  being $p$ and $c$, respectively, the modulus of the EM momentum and the light velocity in vacuum 
of each single mode.
Notice, that the classical-mode energy is a continuous function of the EM momentum and, being independent of the EM frequency, is responsible for a divergent value of the internal total energy, also  known as ultraviolet catastrophe.
\par
By writing the  first and second law of thermodynamics in differential form for a  grand canonical ensemble representing the black-body as:
\begin{equation}
dU=dQ - dL + \mu dN
\end{equation}
with $Q$ and $L$ being, respectively,  the  heat absorbed and  the mechanical work done  by the ensemble, $N$ the number of modes in the system and $\mu$ the associated chemical potential. 
From the interrelations between thermodynamics and the equation of state it is
\begin{equation}
dQ=T \frac{\partial P}{\partial T}dV
\end{equation}
\begin{equation}
dL=PdV 
\end{equation}
\begin{equation}
dU=3 P dV
\end{equation}
and, by assuming with Boltzmann $\mu dN=0$ \cite{kelly81},
it follows the differential equation \cite{montambaux18,reggiani22}
\begin{equation}
\frac{dP}{P} = 4 \frac{dT}{T}
\end{equation}
which solution gives the SB laws for, respectively, $P$ and $U$ as function of $T$:
\begin{equation}
P=\frac{C}{3}T^4
\end{equation}
\begin{equation}
U=CVT^4
\end{equation}
where $C$ is an integration constant keeping the dimensionality of the macroscopic laws for $P$ and $U$ and that, at the time of SB  was determined by comparison with available experiments.  
The validation of the law for $U$ was quite good, and thus the value of 
$C$, also called the radiation constant, was taken as a fundamental constant in classical physics.
The validation by experiments of the law for $P=P(T)$ took some time in view of the low values taken by  $P$, and was credited  to Lebedev  \cite{lebedev1901} and independently to  Nichols and Hull \cite{nichols1903}.
The value of $C$ determined by fit with experiments was:
\begin{equation}
C
= 4 \sigma /c = 7.57 10^{-15} \ J m^{-3 } K_B^{-4} 
\end{equation}
with $\sigma$ the SB constant.
From the above one can  conclude that the SB laws were obtained by Boltzmann on the basis of classical physical concepts, as generally claimed in literature and 
in particular in a recent paper by Jiao Wang, Giulio Casati, and Giuliano Benenti (WCB), with the title {\em Classical Physics and Blackbody Radiation} \cite{wang22}. 
However, as noticed before,  Boltzmann supplemented classical physics with the implicit assumption of a finite value of $U(V,T)$ for given values of $V$ and $T$, 
an assumption not justified by Maxwell equations and, furthermore, the value of the radiation constant could not be determined from classical theories \cite{reggiani22}.
%
\section{Planck model}
With the birth of quantum mechanics, announced by Planck in 1901 \cite{planck01},
the classical normal-modes were replaced by the quantum normal-modes, known as  photons as coined by Lewis in 1926 \cite{lewis26} with energy  depending linearly on frequency and quantized as:
\begin{equation}
\epsilon_p=n hf 
\end{equation}
where the $p$ subscript stands for photons being, respectively, $n$ an integer number, 
$h$ the Planck constant, and $f$ the frequency of the involved photons, with photons obeying the quantum statistics introduced by  Planck. 
Among the predictions of the quantum statistics there were the microscopic derivation of the SB law and the determination of the SB universal constant given in terms 
of universal constants as:
\begin{equation}
\sigma = \frac{2 \pi^5 K_B^4}{15 h^3 c^2}
= 5.67 \times 10^{-8} \ W m^{-2} K^{-4}
\end{equation}
with $K_B$ the Boltzmann constant. 
\par
The presence of Boltzmann and Planck constants together with the light velocity is generally considered to be a signature of the inclusion  of both classical and modern physics,  i.e. relativity and quantum mechanics. 
\par 
Further property of the photon gas inside a black-body is that its instantaneous number is not conserved, and its average number is finite and depends on temperature and volume as 
explicitly given by \cite{reggiani22,leff02}: 
\begin{eqnarray}
\overline N_p(V,T)  &=&  \frac{8 \pi \Gamma(3) \zeta(3) }{c^3 h^3}\ V (K_BT)^3 \nonumber \\ 
 &=&  (2.02 \ 10^7)  \ T^3   \ (m^{-3}) 
\end{eqnarray}
with  $\Gamma(3) \zeta(3) =\int_0^{\infty} x^2/(e^x-1) dx = 2.404$, being $\Gamma$ 
and $\zeta$ respectively the Gamma and the Riemann functions.
\par
In addition, also the average value of the internal-energy is finite and explicitly given by:
\begin{eqnarray}
\overline U_p(V,T) &=&  \frac {8 \pi \Gamma(4) \zeta(4)} {c^3 h^3}  V (K_BT)^4 \nonumber \\ 
&=&  7.57 \ 10^{-16} \ V  T^4 \ (Jm^{-3}) 
\end{eqnarray}
with $\Gamma(4) \zeta(4) = \pi^4  /  15=6.49$.
Accordingly,  one can write:
\begin{equation}
\overline \epsilon_{p} = \frac{\overline U_p} {\overline N_p} =
 \ 2.7 \  K_BT
\end{equation}
with $\overline{\epsilon_{p}}$ being the average energy per photon mode.
\par
We notice that, because of the quantum statistics the numerical value of $2.7$ is slightly less than the value of $3$ pertaining to the classical case  per full relativistic massive particles. 
\par
In particular, the above equation  reminds the property of the classical massive gas where the internal energy is known to depend only from the absolute temperature and can be written as proportional to the product between particles number and single particle average energy. However, for photons the particle number is dependent from both volume and temperature and therefore is not an independent variable, contrary to the massive classical case.
We also remark, that the $T^4$ power law can be partitioned into a  $T^3$ contribution, that is associated with the use of a 3-dimensional geometry, times a $T$ contribution, that is associated with the single particle average-energy. For a 2- or 1-dimensional geometry the exponent value of 4 is expected to reduce accordingly to a value of 3 and 2, as predicted in \cite{alnes07} and found by  the theoretical simulations for the 1-dimensional case \cite{wang22}. 
\par
We want to stress,  that the physical description of the quantum photon-gas 
follows the same scheme used by Boltzmann for the classical EM mode-gas.
Indeed, the internal energy of a photon gas  and the associated state equation take analogous definitions  when the classical $U(V,T)$ is substituted by the quantum $\overline U_p(V,T)$ and in particular a finite value of the quantum average internal-energy as well as of the average  photon-number is naturally implied by the quantum theory  without requiring a supplemented assumption as done by Boltzmann for his classical derivation.  Furthermore, the explicit theoretical value of the radiation constant comes out to agree quite well with the experimental value, without  forcing the fit with experiments. In this way, also the ultraviolet catastrophe is avoided, even if in the classical limit $h \rightarrow 0$ the radiation constant of the quantum  SB law tends to infinite  thus recovering the ultraviolet catastrophe predicted by classical physics.
\section{Conclusions} -
We conclude that, from a macroscopic point of view, Boltzmann classical derivation of the Stefan law by including implicitly the assumption of a finite value for the internal total energy is formally compatible with the Planck quantum derivation. 
Indeed, the quantum property of the photon gas is responsible for the elimination of the divergence of  both the total (average) internal energy and the total (average) number of photons, thus providing the physical justifications for the constraint of a finite total energy implicitly (Boltzmann) and explicitly (WCB \cite{wang22}) made within a classical approach  to the black-body problem.  
In this context, numerical simulations \cite{wang22} including more realistic models could be of interest to evaluate quantitatively the radiation constant for a one-dimensional geometry.
\par
As final remark, the fact that classical simulations of WCB \cite{wang22}  are found compatible with the SB law is here interpreted to be due to the supplemented assumption made by Boltzmann and WCB \cite{wang22}  to take a finite value for the total internal EM energy.
Indeed, this assumption is also sufficient to avoid the ultraviolet catastrophe predicted by
the classical approach of Rayleigh and Jeans \cite{r00,j00}.
However, the quantum features of the SB law stems from the microscopic  value of the  radiation constant $C$, whose first principle derivation contains $h^2$ at the denominator, thus implying that in the classical limit of $h \rightarrow 0$ an infinite value for both the total internal energy and the number of normal modes. 
As a matter of fact, the $T^4$ behavior comes from the assumption of the explicit dependence of $U$ from the volume and the temperature and from the equation of state $U=3PV$, supplemented by the constraint that the internal energy of the black-body is finite. 
These are  the reasons why also the simulations of WCB \cite{wang22} do  not evidence the Rayleigh-Jeans catastrophe \cite{r00,j00}, and thus do not support the research of any  critical reconsideration of the role of classical physics for the understanding of quantum mechanics. 
These considerations were already announced by Planck \cite{planck01}, whose 
quantum theory provides also the solution of the  puzzling concerning the 
classical/quantum compatibility of the SB law by noticing that the implicit assumption made by Boltzmann for a finite value of the total internal energy of the EM gas is fully justified by quantum statistics. In other words, without his knowledge Boltzmann was a 
precursor of Planck quantum statistics.

\end{document}